# Ultrafast, all-optical coherence of molecular electron spins in room-temperature, aqueous solution

Erica Sutcliffe[1], Nathanael P. Kazmierczak[1], Ryan G. Hadt[1]*

[1]Division of Chemistry and Chemical Engineering, Arthur Amos Noyes Laboratory of Chemical Physics, California Institute of Technology; Pasadena, California 91125, United States.

*Corresponding author email: rghadt@caltech.edu

**The tunability and spatial precision of paramagnetic molecules makes them attractive for quantum sensing. However, usual microwave-based detection methods have poor temporal and spatial resolution, and optical methods compatible with room-temperature solutions have remained elusive. Here, we utilize pump-probe polarization spectroscopy to initialize and track electron spin coherence in a molecule. Designed to efficiently couple spins to light, aqueous $K_2IrCl_6$ enables detection of few-picosecond free induction decay at room temperature and micromolar concentrations. Viscosity is found to strongly vary decoherence lifetimes. This work redefines the meaning of room-temperature coherence by improving experimental time resolution by up to five orders of magnitude. Doing so unveils a new regime of electron spin coherence, opening the door to new synthetic design and applications of molecular quantum bits.**

Molecular quantum bits (qubits) are highly tunable and spatially precise, making them very desirable for quantum sensing applications (*1*). By employing molecules with unpaired electron spins, coherent superposition states may be generated between the Zeeman spin sublevels, enabling sensing through distinctly quantum degrees of freedom (*2*). Yet to compete with existing techniques of biological imaging, spin-based quantum sensing must stand up to two key requirements. First, the spin system must display room-temperature coherence in the solution phase – the domain of dynamic biological processes. Second, the system should allow all-optical detection of the spin dynamics to enable precise microscopy. Anionic nitrogen vacancies in diamond ($NV^-$) satisfy both requirements, enabling impressive applications such as single-atom nuclear magnetic resonance (NMR) (*3*), mapping intracellular molecular dynamics (*4*), strain sensing in nanoscale devices (*5*), and imaging magnetic fields in live magnetotactic bacteria (*6*). However, $NV^-$ centers contain an intrinsic spatial constraint of several nanometers owing to the bulk of the lattice; they also have limited chemical tunability. Satisfying both sensing requirements in a molecular system would open a new realm of quantum sensing at the sub-nanometer scale.

Achieving simultaneous room-temperature coherence and all-optical addressability in a molecular qubit system provides significant challenges due to the intrinsic constraints of pulse electron paramagnetic resonance (EPR) and optically detected magnetic resonance (ODMR). These are the two leading spectroscopic techniques for measurements of the electron spin decoherence time of an initialized quantum state, referred to as the spin-spin relaxation time ($T_2$). Both pulse EPR and ODMR require electrical gating of microwave and optical pulses, precluding sub-nanosecond time resolution and thereby placing a significant constraint on which systems can be considered room-temperature coherent. In pulse EPR, short microwave pulses are applied on resonance with transitions between Zeeman-split $M_S$ sublevels – **Fig. 1A**, left. A select few compounds show measurable $T_2$ at room temperature, including Cu(II) and V(IV)O compounds such as vanadyl phthalocyanine (VOPc) (*7*). However, the long wavelength of microwaves limits EPR imaging to a spatial resolution of hundreds of micrometers. ODMR protocols have been described for measuring $T_1$ all-optically (*8*), though coherent state manipulation and measurement of $T_2$ still must use microwave pulses. Detection of an ODMR signal requires unique molecular design principles for spin-selective luminescence. Through mimicking the electronic structure and emission-based detection of the diamond NV$^-$ center, $S = 1$ optically addressable qubits based on Cr(IV) have demonstrated optical spin addressability using ODMR (*8*) – **Fig. 1A**, center. Yet $S = 1$ qubit candidates based on Cr(IV), V(III), or Ni(II) have displayed deleteriously fast spin relaxation rates, with coherence undetectable by EPR above 60 K even in the best cases (*8–11*). Coherence can still be initialized above 60 K, but the spins decohere faster than can be measured. Excited-state spin coherence at room temperature has been reported for organic radical systems using ODMR, but the excited state nature places further constraints on time resolution (*12–14*). Thus, sub-nanosecond measurements of $T_2$ have the potential to unveil a new paradigm for molecular qubits displaying room-temperature coherence.

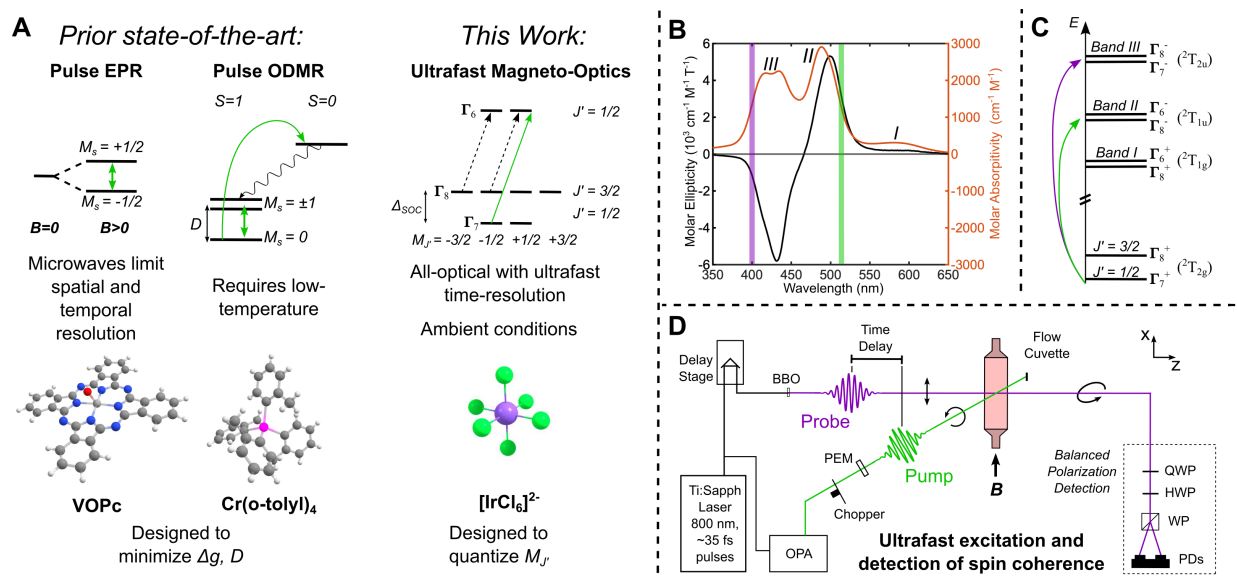

**Fig. 1. Methods for detecting spin coherence.** (**A**) Methods and limitations of spectroscopic techniques for probing electron spin coherence, with characteristic molecules and molecular design criteria. (**B**) MCD and electronic absorption spectra of $K_2IrCl_6$ dissolved in $H_2O$ with peaks labelled and pump/probe wavelengths highlighted. (**C**) Electronic structure of $[IrCl_6]^{2-}$. (**D**) Schematic of the setup used to measure ultrafast spin coherence. The angle between pump and probe pulses is exaggerated for clarity. Abbreviations, and the setup in full, are described in the Supplementary Materials.

Here, we demonstrate picosecond all-optical detection of electron spin coherence in a rationally-designed molecule. Central to our approach is the co-design of ultrafast magneto-optical instrumentation and molecular electronic structure. The decoherence rate can be measured robustly at room temperature in aqueous conditions and at low concentration. Proof-of-concept $T_2$ sensing of solvent viscosity is presented. Comparison to pulse EPR measurements shows the magneto-optical instrumentation can access molecular spin coherence at orders-of-magnitude faster timescales and higher temperatures.

**Design Criteria for All-Optical Molecular Qubits**
The spatial and temporal limits of EPR have been successfully overcome in the semiconductor physics community through the use of ultrafast laser pulses (*15–17*). At the Γ point of a cubic semiconductor, the sublevel scheme shown in **Fig. 1A**, right, is found (*18*). Here, the spin of the electron couples to a threefold orbitally degenerate state (i.e., one with in-state orbital angular momentum, OAM) at the valence band edge, which splits the state into two levels by an amount, $\Delta_{SOC}$, proportional to the magnitude of spin-orbit coupling (SOC). The split states, of symmetry $\Gamma_7$ and $\Gamma_8$, have different total effective angular momentum $J'$, and when their sublevels are eigenstates of $M_{J'}$, circularly polarized light changes the spin state with complete efficiency (*19, 20*). Therefore, selective excitation of either $\Gamma_7 \rightarrow \Gamma_6$ or $\Gamma_8 \rightarrow \Gamma_6$ with circularly polarized light generates a net spin polarization. With a circularly polarized "pump" pulse to initialize the state, the spin coherence is subsequently detected through the change in optical polarization (*15–17*), or even intensity (*19, 21*), of a weaker "probe" pulse. Application of a perpendicular magnetic field causes characteristic Larmor precession of the spin polarization. Such transient magneto-optic methods have temporal resolution of $< 10^{-13}$ s, broad spectral sensitivity (*22–24*) and can be utilized for microscopy with sub-micron spatial resolution (*25, 26*). Despite these clear advantages, whether this technique can be applied effectively to molecular qubits has been an open question.

Molecular systems present two key obstacles to efficient magneto-optical detection. First, unlike cubic semiconductor systems, $J'$ is not a good quantum number in most paramagnetic molecules. In transition metal complexes, this arises from quenching of OAM via the ligand field (*27*). When the sublevels are not quantized by $M_{J'}$, circularly polarized light instead has only a very weak probability to alter the spin state through electric dipole transitions, relying on perturbative out-of-state SOC. Second, the individual $J'$ states must be selectively addressable. While direct-gap semiconductors have narrow excitonic transitions that make addressing only one $J'$ state straightforward, most molecular systems possess much broader electronic transition linewidths. Both challenges may be solved by choosing a high symmetry transition metal complex with a threefold orbitally degenerate ground-state, which maximizes the OAM available to couple to the spin. Octahedral and tetrahedral complexes can produce such ground states according to the Tanabe-Sugano diagrams (*27*). If the SOC is large enough such that $\Delta_{SOC} \gg k_B T$, then the thermally populated ground states will have well-defined $J'$, enabling selective addressability. Because $M_{J'}$ is a good quantum number, circular polarization electric dipole selection rules will be exactly obeyed. Magnetic circular dichroism (MCD) arises from similar microscopic means (*28*), so we expect that suitable candidates also exhibit large MCD signals at the relevant transitions. These criteria contrast with conventional wisdom for the development of molecular qubits exhibiting coherence at elevated temperatures, where one instead seeks to minimize OAM and in turn minimize the orbital contribution to the g value (29), $\Delta g$, and the zero-field splitting, $D$ (*30*) - **Fig. 1A**, bottom.

To the best of our knowledge, time-resolved magneto-optical measurements have so far only been applied to one molecular system: aqueous $CuSO_4$ (*31*). However, the electronic ground state of $D_{4h}$ Cu(II) ions is orbitally nondegenerate, which does not satisfy the requirements for the efficient in-state SOC described above. Therefore, the initialization and readout rely on the much weaker perturbative mixing of excited states into the ground state through out-of-state SOC. Because of this, optical detection of decoherence required prohibitively high Cu(II) concentrations (~1 M) and pulse energies (~10 µJ per pulse). As a consequence, comparatively large nonlinear artifacts such as the optical Kerr effect (OKE) limited the temporal resolution to ~$10^{-10}$ s.

A system satisfying all of these criteria is $K_2IrCl_6$, an air- and water-stable $S = ½$ complex. Its electronic absorption and MCD spectra exhibit three strong bands in the visible (**Fig. 1B**). Extensive prior spectroscopic characterization on this compound revealed the electronic structure shown in **Fig. 1C** (*32–34*). Since the unpaired electron principally resides on the Ir(IV) center (*35, 36*), $\Delta_{SOC}$ is ~5000 cm$^{-1}$ (*37*), giving the ground state $\Gamma_7^+$ symmetry and $J' = ½$ at room temperature. The three bands seen in the spectra correspond to ligand-to-metal charge transfers. Since these result in the electron spin on the ligands, the excited state $\Delta_{SOC}$ is much smaller and, therefore, unresolvable, though this has minimal impact on spin initialization. The large MCD signal (**Fig. 1B**) for the parity-allowed bands II and III further suggests these are ideal candidates to use for initialization and readout.

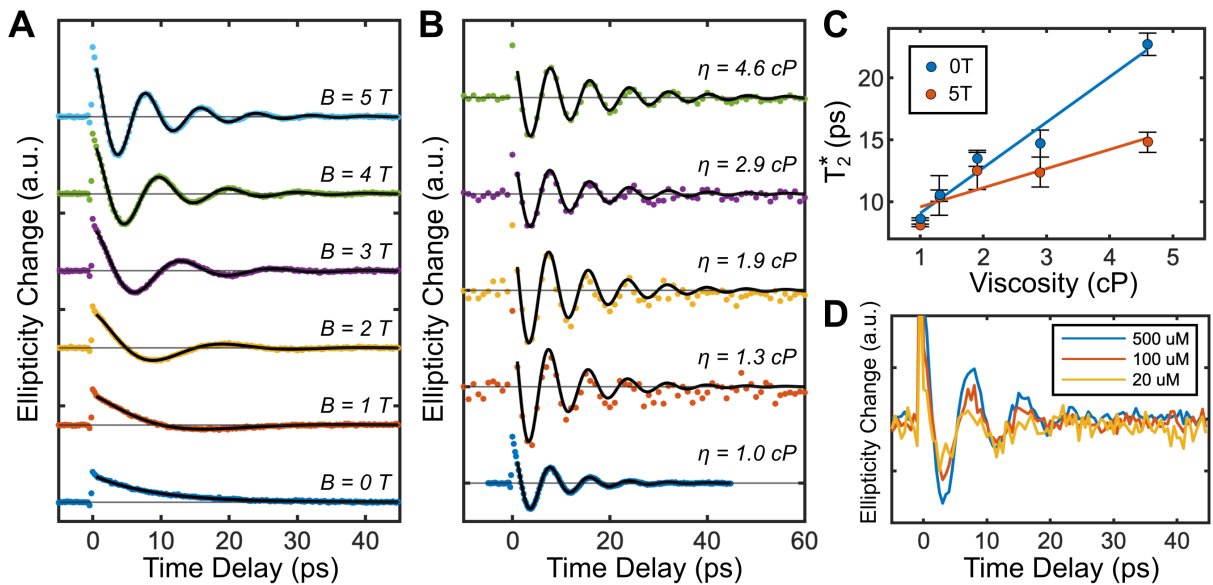

**Fig. 2. Ultrafast free induction decay at room temperature.** (**A**) TRFE measurements on $K_2IrCl_6$ dissolved in H$_2$O at various field strengths with fits to **Eq. 1** shown in black. (**B**) In water:glycerol mixtures to modify viscosity at 5 T with fits to **Eq. 1** shown in black. (**C**) Viscosity dependence of $T_2^*$ measured at 0 T and 5 T alongside corresponding linear fits. (**D**) Concentration sensitivity in H$_2$O at 5 T. OKE spike at time-zero omitted to best show free-induction decay.

**Ultrafast Detection of Free Induction Decay**
To measure electron decoherence, we use the setup shown in **Fig. 1D**, which is described fully in the Supplementary Materials. Spin polarization is initialized along $z$ by a 512 nm, circularly-polarized pump pulse. An almost-collinear, 400 nm probe pulse records the spin polarization along $z$ for a given time-delay as a change in Faraday ellipticity, detected using bridged photodiodes. The sample is flowed to minimize photodegradation and, using a room-temperature-bore superconducting magnet, a field of up to 5 T can be applied along $x$ to induce Larmor precession. A photoelastic modulator (PEM) is used to circularly polarize the 1 µJ pump pulses. The source laser is triggered at 1.014 kHz off of the 50.176 kHz oscillation frequency of the PEM to ensure every consecutive pump pulse has orthogonal polarization (*38*); right- and left-handed circularly polarized pump pulses initialize the spin polarization in opposing directions along $z$. Detection at the frequency of pump intensity and polarization modulation thus gives only the change in probe polarization that is pump-dependent and odd with respect to the initialized polarization. This has the effect of greatly reducing potential OKE artifacts in solution phase measurements. Varying the time-delay between pump and probe yields the time-resolved Faraday ellipticity, TRFE.

Characteristic damped oscillations of free induction decay are observed in the TRFE data for a 2 mM solution of $K_2IrCl_6$ in $H_2O$ (**Fig. 2A**). This decay corresponds to the dephasing of electron spins; it also includes any potential effect of field inhomogeneities on dephasing, so it is termed $T_2^*$, as is commonplace in NMR. The TRFE traces are fit to

$$\eta(t) = \eta_0 \exp(-t/T_2^*) \cos(\omega_L t + \phi) \quad (1)$$

where $\eta_0$, $T_2^*$, and $\phi$ are free parameters; the Larmor frequency, $\omega_L = g_{iso}\mu_B B/\hbar$, provides an isotropic $g$ value, $g_{iso} = 1.74$, with magnetic field strengths, $B = 0 - 5$ T. These fits are shown as black lines in **Fig. 2A** and yield an almost exact description of the data beyond 0.5 ps. These data unambiguously demonstrate the observation of ultrafast free induction decay of the molecular Ir(IV)-based electron spins. At 0 T in $H_2O$, $T_2^*$ is 8.60 ps and spans 9.27 to 8.14 ps across the range of field strengths (**fig. S2**, **table S1**).

Pumping band II and probing the higher energy band III reduces the potential impact of excited-state electronic relaxation on the observed $T_2^*$. Transient absorption spectroscopy was used to verify that $T_2^*$ is not merely determined by excited-state electronic relaxation (**figs. S6-9**). $K_2IrCl_6$ in $H_2O$ exhibits an excited-state lifetime of 17 ps at 0 T, almost double $T_2^*$. Additionally, in DMSO, the excited-state lifetime is 430 ps, yet $T_2^*$ is 17.6 ps (**figs. S3,7**). Together with the good monoexponential fit, these data suggest the excited-state population plays a minimal role in the measured $T_2^*$.

One further, albeit subtle, advantage of this technique over pulse EPR is that the applied magnetic field can be varied continuously across an arbitrarily large range and even eliminated without impacting our ability to measure $T_2^*$. This makes studying the field dependence of $T_2^*$ relatively straightforward, where we observe an overall decrease in $T_2^*$ with $|B|$ (**fig. S2**), likely due to the inhomogeneous $g$ values in the bulk system (*39*). This behavior is mirrored across solvents, with $T_2^*$ in DMSO exhibiting a much stronger field dependence than in $H_2O$ (**fig. S3**). The phases of the oscillations show a roughly $B^3$ dependence, likely due to the optomagnetic field of the pump pulse (Supplementary Text).

### Sensing Viscosity with $T_2^*$

On the picosecond timescale at room temperature, dephasing is expected to involve molecular tumbling contributions (*40*). To test this, 0 and 5 T TRFE traces were recorded for water:glycerol solutions (up to 40% glycerol), which systematically varies the viscosity. At 5 T, $T_2^*$ increases significantly with increasing solution viscosity (**Fig. 2B**). An increasing trend in $T_2^*$ vs. viscosity is observed for both the 0 T and 5 T cases (**Fig. 2C**). At 5 T, the trend is complicated by the oscillations and inhomogeneities in $g$ value. However, for 0 T, $T_2^*$ varies linearly with viscosity and more than doubles in magnitude (8.60 and 21.9 ps in neat $H_2O$ and 3:2 water:glycerol, respectively). Since the molecular tumbling time is also proportional to viscosity (*41*), such a trend is consistent with this dephasing mechanistic hypothesis. These observations suggest that complex immobilization in biological macromolecules should significantly prolong coherence times. Hyperfine coupling to solvent molecules can also contribute to dephasing along with molecular tumbling; indeed, dissolving the Ir(IV) complex in $D_2O$ instead of $H_2O$ increases $T_2^*$ from 8.60 to 10.1 ps at 0 T (**fig. S5 and table S1**). Future studies of the temperature dependence of $T_2^*$ may additionally elucidate spin-vibrational coupling contributions to decoherence (*42*).

For quantum sensing applications, sensitive detection is vital, and the high sensitivity of EPR is a key reason for its widespread applicability. As shown in **Fig. 2D**, oscillations at 5 T in $H_2O$ can still be observed upon reducing the concentration 100-fold to 20 μM (and using 2 μJ pump pulses), a similar detection limit to EPR (*43*).

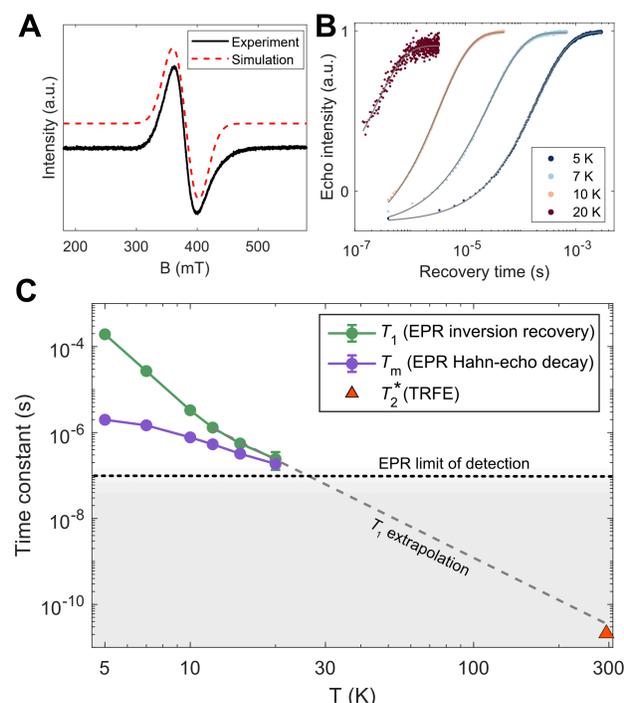

**Fig. 3. Comparison between EPR and TRFE capabilities.** (**A**) X-band CW-EPR spectrum of 2 mM $K_2IrCl_6$ in a 3:2 water:glycerol glass at 15 K (ν = 9.637 GHz). Simulated trace with isotropic *g* = 1.807. (**B**) Q-band pulse EPR inversion recovery traces collected at the maximum microwave echo intensity (1318 mT, ν = 34.110 GHz). (**C**) Comparison between $T_1/T_m$ from pulse EPR and the TRFE measurement; error bars are mostly smaller than data points. Extrapolation line produced from a linear fit to $T_1$ between 12 and 20 K.

**Comparison with EPR**
Pulse EPR currently constitutes the standard method for measuring electron spin coherence lifetimes. We compared the TRFE measurement to the information obtainable from pulse EPR on the same complex (**Fig. 3**). $[IrCl_6]^{2-}$ is an $S = ½$ species with a broad, isotropic continuous wave (CW) EPR signal centered at *g* = 1.807 in a frozen solvent glass (**Fig. 3A**). The 0.07 discrepancy between the *g* values from CW-EPR and TRFE is likely due to the phase transition and wide temperature difference between the measurements. Q-band pulse EPR inversion recovery and Hahn-echo decay measurements were conducted to obtain $T_1$ and $T_m$, respectively. $T_m$ denotes the phase-memory time when applying a single refocusing π pulse for echo detection, and constitutes a common approximation to $T_2$ measured by EPR. Strong echo signals leading to precise time constant determination are observed over a 5 – 15 K temperature range. However, the echo intensity at 20 K becomes very weak owing to fast spin relaxation, leading to degraded signal-to-noise in the inversion recovery trace (**Fig. 3B**). The polarization inversion observed at short time is reduced by 50% relative to the 5 K measurement, indicating a substantial amount of the polarization relaxes during the spectrometer deadtime; this constitutes a $\geq$ 120 ns period after the refocusing pulse during which the spin echo cannot be measured. Due to the deadtime constraint and the minimum pulse timing increment of 2 ns, we estimate the EPR limit of detection for $T_1$ and $T_m$ to be on the order of 100 ns. $T_1$ and $T_m$ could not be reliably ascertained by pulse EPR above 20 K.

The fitted $T_1$ values vary strongly as a function of temperature due to thermal population of vibrational modes that couple to the spin (**Fig. 3C**). By contrast, $T_m$ displays a weaker scaling with temperature below 10 K, which becomes stronger above 10 K as $T_1$ approaches $T_m$. This arises because the maximum value of $T_m$ (the transverse relaxation) is limited by $T_1$ (the longitudinal relaxation), as longitudinal recovery of equilibrium spin polarization necessarily removes magnetization from the transverse plane of the Bloch sphere. Thus, $T_1$ constitutes the fundamental limitation on the observation of high-temperature spin coherence (*40*). We extrapolated $T_1$ beyond the EPR limit of detection for comparison to the TRFE measurements. We find that $T_2^*$ at room temperature and 1.3 T agrees well with the extrapolated prediction for $T_1$-limited-$T_2$ (**Fig. 3C**). This extrapolation should be taken loosely, as it cannot account for the change in relaxation dynamics induced by melting of the water:glycerol glass. However, it suggests overall consistency between the time constants measured by pulse EPR and those measured by TRFE.

In summary, the TRFE magneto-optical approach has detected a spin coherence lifetime that is four orders of magnitude smaller than the pulse EPR limit of detection (8 ps vs. ~100 ns). This extends the temperature range of coherence detection by a factor of 15 for $K_2IrCl_6$, from 20 K to 294 K (**Fig. 3C**). These results demonstrate that TRFE has already accessed a new regime of ultrafast molecular spin dynamics.

**Discussion**
The present work has enabled room-temperature, all-optical quantum sensing by redefining the concept of "room-temperature coherence". This categorization is entirely relative to the timescale of the instrumentation employed to detect the spin coherence. The literature of microwave-addressable qubits has classified compounds as room-temperature coherent or not based on whether they possess spin echo lifetimes beyond the 100 ns limit of detection of pulse EPR, and ligand field design strategies have been described to enable this functionality (*44, 45*). Yet under the ~1 ps limit of detection of the TRFE measurement, molecules not formerly considered "room-temperature coherent" in pulse EPR can now become room-temperature coherent, as exemplified by $K_2IrCl_6$. The fast spin relaxation of $K_2IrCl_6$ can thus be tolerated in order to reap the benefits of $J'$ quantization for polarized optical spectroscopy.

Furthermore, the TRFE scheme transforms "microwave-addressable qubits" into "optically-addressable qubits" by imparting an optical spin interface to molecules with an isolated doublet ground state ($J' = \frac{1}{2}$). This eliminates the requirement to match the non-Kramers NV$^-$ electronic structure for optical addressability. Freed from strict imitation, synthetic efforts in molecular quantum information science may now focus on different design criteria, including air-/water-

/photo-stability, biological compatibility, and maximized ellipticity signal produced through in-state OAM. $K_2IrCl_6$ constitutes the first example following many of these criteria, but we anticipate that further synthetic efforts contain great potential to optimize TRFE molecular quantum sensor design. We additionally believe the TRFE instrumentation is amenable to microscopy implementations with improved spatial precision over EPR imaging. The co-design of magneto-optical instrumentation and molecular sensors constitutes an exciting new path forward for quantum information science.

**Acknowledgments**:

**Funding**: N.P.K. acknowledges support by the Hertz Fellowship and the National Science Foundation Graduate Research Fellowship under Grant No. DGE-1745301. Financial support from the U.S. Department of Energy (DOE), Office of Science, Office of Basic Energy Sciences, Atomic, Molecular, and Optical Sciences program (DE-SC0022920) is gratefully acknowledged.

**Author Contributions**: E. S. and N. P. K. performed the measurements and analyzed the data. R. G. H. advised on all efforts. All authors contributed to conception of the research, as well as composing the manuscript.

**Competing Interests**: The authors have no competing interests to declare.

**Data and Materials Availability**: All study data are included in the article and/or Supplementary Materials and are available upon reasonable request to the corresponding author.